\documentclass[review]{elsarticle}

\usepackage{lineno,hyperref}
\usepackage{bm,color,transparent,amsmath,amssymb,amsthm,fancyhdr,mathrsfs,graphicx,setspace}

\modulolinenumbers[5]

\journal{Composite Structures}

\begin{document}

\begin{frontmatter}

\title{Three-dimensional nonlinear micro/meso-mechanical response of the fibre-reinforced polymer composites}

\author[label1]{Z. Ullah\fnref{label2}}
\author[label1]{{\L}. Kaczmarczyk}
\author[label1]{C. J. Pearce}
\address[label1]{School of Engineering, Rankine Building, The University of Glasgow, Glasgow, G12 8LT, UK}
\fntext[label2]{Correspondence to: Z. Ullah, E-mail: Zahur.Ullah@glasgow.ac.uk}

\begin{abstract}
A three-dimensional multi-scale computational homogenisation framework is developed for the prediction of nonlinear micro/meso-mechanical response of the fibre-reinforced polymer (FRP) composites. Two dominant damage mechanisms, i.e. matrix elasto-plastic response and fibre-matrix decohesion are considered and modelled using a non-associative pressure dependent paraboloidal yield criterion and cohesive interface elements respectively. A linear-elastic transversely isotropic material model is used to model yarns/fibres within the representative volume element (RVE). A unified approach is used to impose the RVE boundary conditions, which allows convenient switching between linear displacement, uniform traction and periodic boundary conditions. The computational model is implemented within the framework of the hierarchic finite element, which permits the use of arbitrary orders of approximation. Furthermore, the computational framework is designed to take advantage of distributed memory high-performance computing. The accuracy and performance of the computational framework are demonstrated with a variety of numerical examples, including unidirectional FRP composite, a composite comprising a multi-fibre and multi-layer RVE, with randomly generated fibres, and a single layered plain weave textile composite. Results are validated against the reference experimental/numerical results from the literature. The computational framework is also used to study the effect of matrix and fibre-matrix interfaces properties on the homogenised stress-strain responses. 
\end{abstract}


\end{frontmatter}


\section{Introduction}\label{sec_intro}

Compared to conventional materials, fibre-reinforced polymer (FRP) composites can offer exceptional physical and chemical properties (including high strength, low specific weight, fatigue and corrosion resistance, low thermal expansion and high dimension stability), making them ideal for a variety of engineering applications, including aerospace, marine, automotive industry, civil structures and prosthetics \cite{Tong2002, Mouritz1999, ZUllah_CAS_2016}. Phenomenological or macro-level models cannot accurately describe the complex behaviour of FRP composites due to their underlying complicated and heterogeneous microstructure. Furthermore, nonlinearities associated with the matrix elasto-plasticity and fibre-matrix decohesion make the computational modelling even more challenging. Multi-scale computational homogenisation (CH) provides an accurate modelling framework to simulate the behaviour of FRP composites and determine the macro-scale homogenised (or effective) response, based on the physics of an underlying, microscopically heterogeneous, representative volume element (RVE) \cite{NematNaseer1993, Geers2010, Xiaoyi2016_textile, UllahACME2014, UllahACME2015, UllahACME2016, ZUllah_CAS_2016}. The homogenised properties calculated from the multi-scale CH are subsequently used in the numerical analysis of the macro-level structures. 

A variety of numerical techniques have been developed to model the nonlinear micro-mechanical response of unidirectional (UD) FRP composites, mostly based on finite element analysis. For UD glass/carbon (G/C) FRP composites, a computational model was developed in \cite{Gonzalez2007, NayaLLorca2015_incollection} within the framework of finite deformation. Both in-plane shear and compressive loading scenarios were considered. The Mohr-Coulomb yield criterion and cohesive interface elements were used to model the response of epoxy matrix and fibre-matrix interfacial decohesion respectively. Fibres were generated randomly within the RVEs using the algorithm presented in \cite{Brockenbrough1991} and were modelled as a linear-elastic and isotopic material. A parametric study, including the effect of matrix and interface properties on the stress-strain response, was also conducted. The idea of \cite{Gonzalez2007} was extended further in \cite{Vaughan2011} by incorporating thermal residual stresses (due to cooling of FRP composites after curing process, caused by the mismatch in thermal expansion coefficients of matrix and fibres) in the simulation, in addition to transverse tensile and cyclic loading for the CFRP composites. The nearest neighbour algorithm (NNA) \cite{Vaughan2010} developed by the same authors, was used to randomly generate the fibres within the RVEs. Using the same constitutive models for matrix, fibres and fibre-matrix decohesion as in \cite{Gonzalez2007}, a multi-layer multi-fibre (M$^2$) RVE was used in \cite{Soni2014_M2RVE} for laminates. Each lamina was modelled as a cube with randomly distributed but axially aligned fibres, generated using a fibres randomisation algorithm in DIGIMAT FE \cite{Digimat2011}. Both cross [0/90]$_{ns}$ and angle [$\pm$45]$_{ns}$ (where the subscript $ns$ represents $n$ layers with the same sequence and symmetric about the mid plane) GFRP composites were considered with in-plane shear loading and results of stress-strain behaviour were validated against the experimental results. A combined transverse compression and axial tension loading scenario was considered in \cite{Romanowicz2012} for UD GFRP composite. In addition to matrix plasticity and fibre-matrix decohesion, fibre breakage was also included in the FE simulation. The pressure dependent, Drucker-Prager yield criterion was used to model matrix plasticity and both fibre breakage and fibre-matrix interfacial decohesion were modelled with cohesive interface elements. A simple periodic, hexagonal fibre arrangement was assumed. In \cite{Parambil2015}, a modified von Mises yield criterion was used to model the behaviour of the matrix material, while a maximum tensile stress criterion was used to model fibre breakage. Fibre-matrix decohesion was also included in the simulation and was modelled with cohesive interface elements. The random distribution of the fibres was also included within the RVE based on the optical microscopy of real composites. A variety of loading conditions was used subsequently to study the response of the UD FRP composite. The limitations of different plasticity models for modelling matrix materials, including Mohr-Coulomb and Drucker-Prager were argued in \cite {Melro2013_partI, Melro2013_partII}, especially in complex loading scenarios. Instead of the conventional plasticity models, a pressure dependent thermodynamically consistent plasticity model \cite{Tschoegl1971} was used. A statistically proven random distribution algorithm proposed by the same authors in \cite{Melro2008_rado_fibrer} was used to randomly generate UD fibres within the RVEs. Similar to previous studies, fibres were modelled as linear-elastic and isotropic material and fibre-matrix decohesion was modelled with cohesive interface elements. A variety of RVE loading scenarios was considered including transverse tension and compression, transverse and longitudinal shear and combined transverse compression and transverse shear. 

A number of numerical modelling approaches have been used to simulate the behaviour of textile composites subjected to different loading scenarios. A comprehensive review of these methods can be found in \cite{Matveev_Long_2015_incollection}. Continuum damage mechanics (CDM) was used in \cite{kollegal2001_textile} to model both matrix and yarns for glass and carbon plain weave textile composites. Dissipated energy density was used as damage parameter and both material and geometric nonlinearities were included in the simulation. Further use of CDM in the simulation of textile composites can also be found in \cite{ZakoTextileCDM2003, Blackketter1993, Ivanov2009}. Moreover, a three-dimensional CDM based approach was used to simulate the progressive damage in laminated FRP composites in \cite{Falzon2011CDM_2011_PartI, Falzon2011CDM_2011_PartII}. A variety of failure mode, including matrix tensile and compressive cracking, fibre tensile and compressive failure, fibre-matrix shearing and delimitation between the layers were included in the simulations. For a twill weave textile CFRP subjected to in-plane loading, a meso-mechanical analysis was performed in \cite{Stier2015}. The matrix was modelled as elasto-plastic material with the same plasticity model as in \cite{Melro2013_partI, Melro2013_partII, Tschoegl1971}, while yarns were modelled as linear-elastic and transversely isotropic material. Results of the RVE strain fields and homogenised stress-strain response were validated against the experimental results and found in a good agreement. 

These numerical simulations, described above, of FRP composite behaviour are limited to specific RVE type (2D or 3D, UD or woven/textile) or loading scenarios (normal or shear). In contrast, this paper develops a fully generalised three-dimensional micro/meso-mechanical framework, which is subsequently implemented in the authors' open source FE software, MOFEM \cite{MoFEM_2016}. The dominant damage mechanisms (observed experimentally \cite{Gonzalez2007}), i.e. matrix elasto-plasticity and fibre-matrix decohesion, are included in the computational framework. Matrix material is modelled using a pressure dependent paraboloidal yield criterion \cite {Melro2013_partI, Melro2013_partII, Stier2015, Tschoegl1971} with an exponential hardening law. Fibre-matrix decohesion is modelled with zero thickness cohesive interface elements. Yarns are modelled as linear-elastic and transversely isotropic materials. Rather than simplified fibre arrangements for UD FRP composites, e.g. in \cite{whitcomb2002, hobbiebrunken2006, Romanowicz2012, Maligno2008}, which are not the actual representation of the real FRP composites and can lead to erroneous results. this study adopts a statistically proven random distribution algorithm proposed in \cite{Melro2008_rado_fibrer} to generate fibre arrangements within the RVE. The RVE boundary conditions are imposed in a unified manner which allows convenient switching between displacement, traction and periodic boundary conditions \cite{Lukasz2008}. Hierarchic finite elements are adopted, which permits the use of arbitrary order of approximation, leading to accurate results for relatively coarse meshes. The computational framework is designed to take advantage of distributed memory high-performance computing. Moreover, CUBIT \cite{Cubit} and ParaView \cite{Paraview} are used as pre- and post-processor respectively. 

This paper is organised as follows. The computational framework is fully described in \S \ref{sec_com_framework}. The material models are given in \S \ref{sec_mat_model}, consisting of material model for matrix \S \ref{sec_matirx}, yarns/fibres \ref{sec_yarns_fibres} and fibre-matrix interfaces \S \ref{sec_interace}. The nonlinear multi-scale CH with corresponding RVE boundary conditions are explained in \S \ref{sec_compHomo}. Calibration and validation of the matrix plasticity model is given in \S \ref{sec_calib_valid}. Three numerical examples are given in \S \ref{sec_num_examples}, including UD GFRP composites \S \ref{sec_Exp_UGFRP}, M$^2$RVE \S \ref{sec_Exp_M2RVE} and plain weave textile composites \S \ref{sec_Exp_textile}. Finally, the concluding remarks are given in \S \ref{sec_conc_remarks}.

\section{Computational framework}\label{sec_com_framework}
The computational framework developed for FRP composites consists of a set of constitutive models for individual components including matrix, yarns/fibres and fibre-matrix interface and implemented within the formulation of first-order multi-scale CH. 

\subsection{Material constitutive models}\label{sec_mat_model}
\begin{figure}[h!]
\begin{centering}
\includegraphics[scale=0.8]{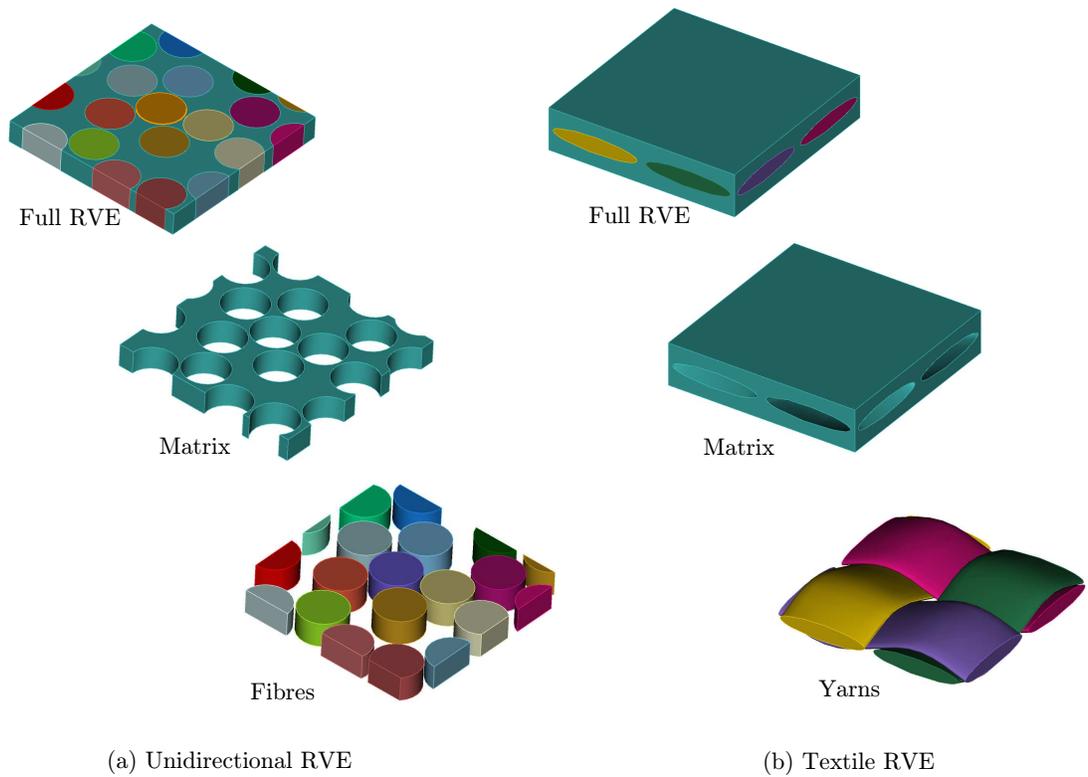}
\caption{Typical RVEs for unidirectional FRP and textile composites} \label{Fig_RVEs}
\end{centering}
\end{figure}

Typical RVEs in the case of UD FRP and textile composites are shown in Figure \ref{Fig_RVEs}(a) and (b) respectively, consisting of yarns/fibres embedded within a polymer matrix. The constitutive model for FRP composites is a combination of constitutive models for these individual components, together with fibre-matrix interface decohesion. In the following, each of these constitutive model is explained in detail. 

\subsubsection{Matrix}\label{sec_matirx}
\begin{figure}[h!]
\begin{centering}
\includegraphics[scale=0.8]{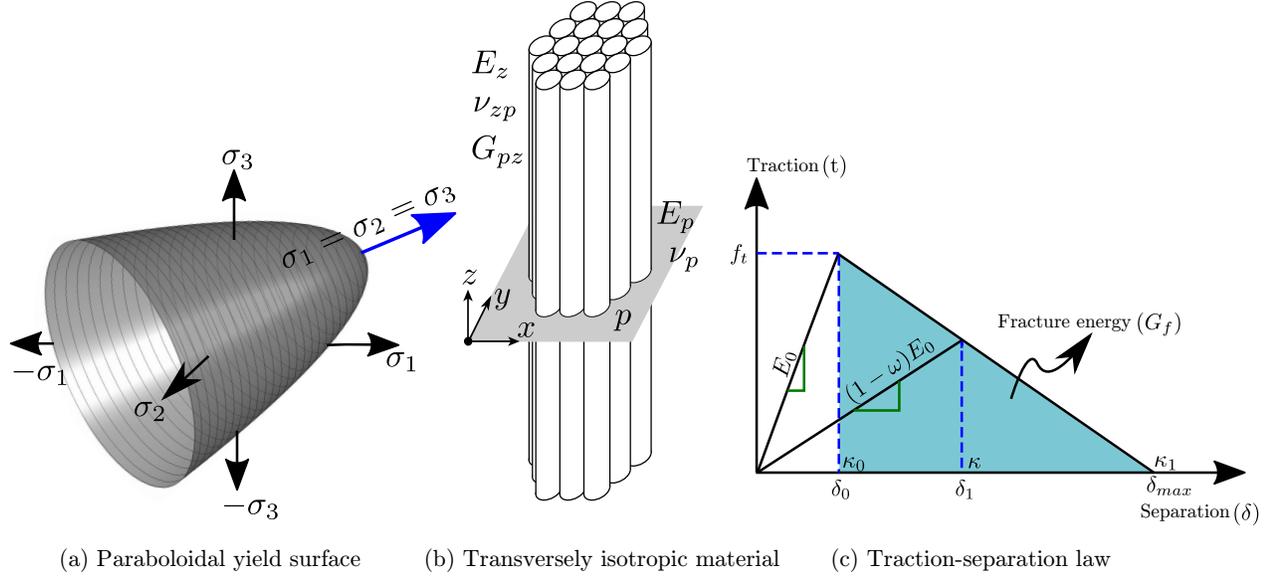}
\caption{Material models for matrix and fibre-matrix decohesion} \label{Fig_mat_model}
\end{centering}
\end{figure}

The polymer matrix is modelled as an elasto-plastic material using a non-associative pressure dependent paraboloidal yield criterion \cite {Tschoegl1971, Melro2013_partI, Melro2013_partII, Stier2015, UllahACME2016}. This plasticity model can incorporate different yield strengths in tension and compression and is shown Figure \ref{Fig_mat_model}(a) in the principal stress space. The yield function is written as
\begin{equation}\label{eq_f}
f\left(\boldsymbol{\sigma},\sigma_{c},\sigma_{t}\right)=6J_{2}+2I_{1}\left(\sigma_{c}-\sigma_{t}\right)-2\sigma_{c}\sigma_{t}, 
\end{equation}
where $\boldsymbol{\sigma}$ is Cauchy stress tensor, $I_1=\textrm{tr}(\boldsymbol{\sigma})$ is the first invariant of Cauchy stress tensor, $J_{2}=\frac{1}{2}\boldsymbol{\eta}:\boldsymbol{\eta}$ is the second invariant of deviatoric stress $\boldsymbol{\eta}=\boldsymbol{\sigma}-\frac{1}{3}I_{1}$ and $\sigma_t$ and $\sigma_c$ are yield strengths in tension and compression respectively. A non-associative flow rule is used, for which the plastic potential function is written as 
\begin{equation}\label{eq_g}
g\left(\boldsymbol{\sigma},\sigma_{c},\sigma_{t}\right)=6J_{2}+2\alpha I_{1}\left(\sigma_{c}-\sigma_{t}\right)-2\sigma_{c}\sigma_{t}, \qquad   \alpha=\frac{1-2\nu_{plas}}{1+\nu_{plas}},  
\end{equation}
where $\nu_{plas}$ is a material parameter and is known as plastic Poisson's ratio. Furthermore, the Helmholtz free energy in the case of linear isotropic hardening is written as
\begin{equation}\label{eq_energy_linear}
\psi=\frac{1}{2}\lambda\textrm{tr}[\boldsymbol{\varepsilon}]^{2}+\mu\boldsymbol{\varepsilon}:\boldsymbol{\varepsilon}+\sigma_{t_{0}}\alpha_{0}+\frac{1}{2}H_{t}\alpha_{0}^{2}+\sigma_{c_{0}}\alpha_{1}+\frac{1}{2}H_{c}\alpha_{1}^{2},
\end{equation}
where $\lambda$ and $\mu$ are the Lame parameters, $\boldsymbol{\varepsilon}$ is the strain tensor, $\sigma_{t_{0}}$ and $\sigma_{c_{0}}$ are the initial yield strengths in tension and compression respectively, $\alpha_{0}$ and $\alpha_{1}$ are internal kinematic variables and $H_t$ and $H_c$ are hardening parameters in case of tension and compression respectively. Following Equation (\ref{eq_energy_linear}), yield strengths in tension and compression are written as
\begin{equation}\label{eq_hardening_linear}
\sigma_{t}=\frac{\partial\psi}{\partial\alpha_{0}}=\sigma_{t_{o}}+\alpha_{0}H_{t},\qquad\sigma_{c}=\frac{\partial\psi}{\partial\alpha_{1}}=\sigma_{c_{o}}+\alpha_{1}H_{c}
\end{equation}
A more realistic, exponential hardening law is presented in this paper, due to which Equation (\ref{eq_energy_linear}) is rewritten as 
\begin{equation}\label{eq_energy_exp}
\psi=\frac{1}{2}\lambda\textrm{tr}[\boldsymbol{\varepsilon}]^{2}+\mu\boldsymbol{\varepsilon}:\boldsymbol{\varepsilon}+\left(\sigma_{t_{0}}+H_{t}\right)\alpha_{0}+\frac{H_{t}}{n_{t}}e^{-n_{t}\alpha_{0}}+\left(\sigma_{c_{0}}+H_{c}\right)\alpha_{1}+\frac{H_{c}}{n_{c}}e^{-n_{c}\alpha_{1}},
\end{equation}   
where $H_t$, $H_c$ are the difference between the ultimate and yield strengths, $n_t$ and $n_c$ are material parameters and determine the rate of convergence between yield and ultimate strengths. Following Equation (\ref{eq_energy_exp}), yield strengths in tension and compression are written as
\begin{equation}\label{eq_hardening_exp}
\sigma_{t}=\frac{\partial\psi}{\partial\alpha_{0}}=\sigma_{t_{o}}+H_{t}\left(1-e^{-n_{t}\alpha_{0}}\right),\qquad\sigma_{c}=\frac{\partial\psi}{\partial\alpha_{1}}=\sigma_{c_{o}}+H_{c}\left(1-e^{-n_{c}\alpha_{1}}\right).
\end{equation}   
A similar hardening law as a function of equivalent plastic strain was also used in \cite{Stier2015}. 

\subsubsection{Yarns/fibres}\label{sec_yarns_fibres}
In textile composites, as shown in Figure \ref{Fig_RVEs}(b), yarns are interwoven together to form a textile structure. On the micro-level, yarns are the same as UD FRP composites consisting of bundles of glass/carbon fibres within the polymer matrix. On the meso-level yarns are modelled as homogenous and transversely isotropic material with homogenised or effective properties obtained from the multi-scale CH of the UD FRP composites. Five material parameters are required for transversely isotropic materials, which are $E_p, \nu_p, E_z, \nu_{pz}$ and $G_{zp}$, where $z$ and $p$ are fibres and transverse directions respectively as shown in Figure \ref{Fig_mat_model}(b). $E_p$ and $\nu_p$ are Young's modulus and Poisson's ratio in the transverse direction respectively, while $E_z$, $\nu_{pz}$ and $G_{zp}$ are the Young's modulus, Poisson's ratio and shear modulus in the fibre directions respectively. In order to re-orient the known stiffness matrix for the transversely isotropic material from the local coordinates to global coordinates, the yarns directions at each Gauss point need to be determined. To do this it is possible to simply use the cubic splines that were used to construct the yarns. However, this can lead to inaccuracies in the case of yarns with non-uniform cross-sections along their length. An alternative approach is used in this paper, in which the yarns directions are determined by solving the potential flow along these yarns. A detailed description of this approach and how to transfer the stiffness matrix from the local to global coordinate axes is given in \cite{Xiaoyi2016_textile, ZUllah_CAS_2016, UllahACME2014, UllahACME2015, UllahACME2016}. 

On the micro-level in the case of UD FRP composites, fibres are modelled as linear-elastic and isotropic material (a special case of transversely isotropic material model), for which only two material parameters, i.e. Young's modulus and Poisson's ratio ($E_f$, and $\nu_f$) are required. In this case, the following material parameters are used: 
\begin{equation}   
E_p=E_z=E_f, \qquad \nu_p=\nu_{pz}=\nu_f, \qquad G_{p}=\frac{E_f}{2\left(1+\nu_f \right)}.
\end{equation}

\subsubsection{Fibres/matrix interfaces}\label{sec_interace}
Fibre-matrix decohesion is modelled using standard cohesive interface elements with a straightforward material model, i.e. linear traction-separation law (shown in Figure \ref{Fig_mat_model}(c)). Only three material parameters are required for the material model, including cohesive strength $f_t$, fracture energy $G_f$ (shaded area in Figure \ref{Fig_mat_model}(c)) and material parameter $\beta$, which assign different weight to opening and shear displacements. Mathematically the material model for cohesive interface elements is written as   
\begin{equation}\label{eq_trac_sep}
t=\left\{ \begin{array}{ll}
 E_{0}\delta & \textrm{if}\,\,\,\,\delta<\delta_{0},\\
 \left(1-\omega\right)E_{0}\delta & \textrm{if}\,\,\,\,\delta_{0}\leq\delta<\delta_{\textrm{max}},\\
 0 & \textrm{if} \,\,\,\,\delta<\delta_{\textrm{max}},
 \end{array}\right.
\end{equation}   
where $E_0$ is the initial stiffness, $\delta=\sqrt{\delta_n^2 + \beta(\delta_{s1}^2+ \delta_{s2}^2)}$ is the displacement jump with $\delta_n$ and $\delta_s$ as its normal and shear components and $\omega$ is damage parameter. $\kappa$ is a history parameter and is equal to the highest value of displacement jump $\delta$. Furthermore, $\delta_0$ and $\delta_{max}$ are respectively the displacement jumps at the onset of damage ($\omega=0$) and when the interface is fully damaged ($\omega=1$). $E_0$ and $\delta_0$ are written as 
\begin{equation}
E_0=\frac{E_m}{h}, \qquad  \delta_0=\frac{f_t}{E_0},
\end{equation}   
where $E_m$ is the Young's modulus of matrix material and $h$ is the interface thickness. Furthermore, from Figure \ref{Fig_mat_model}(c), the damage parameter $\omega$ is written as 
\begin{equation}
\omega=\frac{\left(2G_fE_0 + f_t^2\right)\kappa} {2G_f\left(f_t + \kappa E_0\right)}.
\end{equation}   
The constitutive matrices for cohesive interface elements in the local and global coordinate systems are written as 
\begin{equation}
\textbf{D}^{\textrm{loc}}=\left(1-\omega\right) \textbf{I}E_0,  \qquad  \textbf{D}^{\textrm{glob}}=\textbf{R}^{\textrm{T}} \textbf{D}^{\textrm{loc}} \textbf{R},
\end{equation}   
where $\textbf{I}$ is the unit matrix and $\textbf{R}$ is the transformation matrix \cite{Segurado2004_cohesiveInterface}. Equations for stiffness matrix and corresponding internal forces are written as 
\begin{equation}
\textbf{K}^\textrm{el} =\int_A \boldsymbol{\Phi}^\textrm{T}  \textbf{D}^{\textrm{glob}}  \boldsymbol{\Phi} dA, \qquad \textbf{F}^\textrm{el}_\textrm{int}=\int_A  \boldsymbol{\Phi}^\textrm{T} \textbf{R}^{\textrm{T}} \textbf{t}^{\textrm{loc}}  dA, 
\end{equation}   
where $\boldsymbol{\Phi}$ is a matrix of shape functions and $\textbf{t}^{\textrm{loc}}$ is the traction vector in local coordinates, details of which are given in \cite{Segurado2004_cohesiveInterface}. 

\subsection{Multi-scale computational homogenisation}\label{sec_compHomo}
\begin{figure}[h!]
\begin{centering}
\includegraphics[scale=0.65]{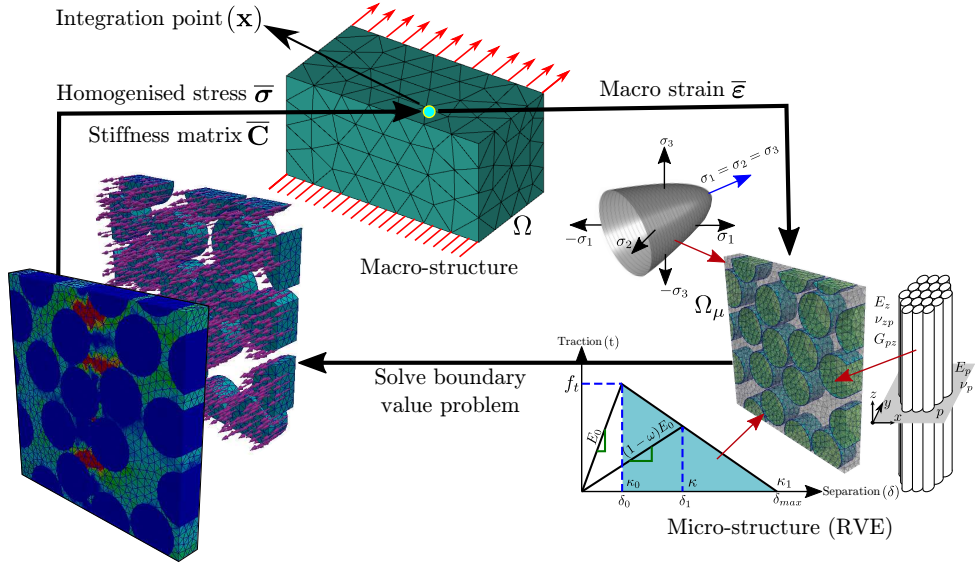}
\caption{Multi-scale computational homogenisation} \label{Fig_multi_scale_CH}
\end{centering}
\end{figure}

In multi-scale CH, a heterogeneous RVE is associated with each Gauss point of the macro-homogeneous structure, the boundary conditions for which are implemented using the generalised procedure proposed in \cite{ZUllah_CAS_2016, Lukasz2008}. Small displacement and small strain formulations are used within the framework of first order multi-scale CH, the basic concept of which is shown in Figure \ref{Fig_multi_scale_CH}, where $\Omega\subset\mathbb{R}^{3}$ and $\Omega_{\mu}\subset\mathbb{R}^{3}$ are macro and micro domains respectively. Macro-strain $\overline{\boldsymbol{\varepsilon}}=\left[\begin{array}{cccccc}\overline{\varepsilon}_{11} & \overline{\varepsilon}_{22} & \overline{\varepsilon}_{33} & 2\overline{\varepsilon}_{12} & 2\overline{\varepsilon}_{23} & 2\overline{\varepsilon}_{31}\end{array}\right]^{T}$ is first calculated at each Gauss point $\mathbf{x}=\left[\begin{array}{ccc}x_{1} & x_{2} & x_{3}\end{array}\right]^T$ of the macro-structure, which is then used to formulate the boundary value problem on the micro-level. After solution of the micro-level boundary value problem, homogenised stress $\overline{\boldsymbol{\sigma}}=\left[\begin{array}{cccccc}\overline{\sigma}_{11} & \overline{\sigma}_{22} & \overline{\sigma}_{33} & \overline{\sigma}_{12} & \overline{\sigma}_{23} & \overline{\sigma}_{31}\end{array}\right]^{T} $ and stiffness matrix $\overline{\mathbf{C}}$ are calculated.

For a global step $n+1$, the discretised system of equations in case of an iteration $i$ of the Newton-Raphson algorithm is written as
\begin{equation}\label{eq_RVE_disc_Eqs}
\left[\begin{array}{ll}
\mathbf{K}_{n+1}^{i} & \mathbf{C}^{T}\\
\mathbf{C} & \mathbf{0}
\end{array}\right]\left\{ \begin{array}{c}
\triangle\mathbf{u}_{n+1}^{i}\\
\triangle\boldsymbol{\lambda}_{n+1}^{i}
\end{array}\right\} =\mathbf{F}_{n+1}^{i}, 
\end{equation}
where $\mathbf{K}$ and $\mathbf{u}$ are the standard FE tangent stiffness matrix and displacement vector respectively and $\boldsymbol{\lambda}$ is the unknown vector of Lagrange multipliers required to impose the RVE boundary conditions. Matrix $\mathbf{C}$ in Equations (\ref{eq_RVE_disc_Eqs}) are calculated over the boundary $\Gamma$ of the RVE and are constant throughout the calculations \cite{ZUllah_CAS_2016, Lukasz2008} and are given as
\begin{equation}\label{eq_mat_C}
\mathbf{C}=\int_{\partial\Omega_{\mu}}\mathbf{HN}^{T}\mathbf{N}d\partial\Omega_{\mu}.
\end{equation}
In Equation (\ref{eq_mat_C}), $\mathbf{N}$ is a matrix of shape functions and $\mathbf{H}$ is a matrix that is specific to the type of boundary conditions used, each row of which represents an admissible distribution of nodal traction forces on the RVE boundary \cite{Lukasz2008}. The specific choice of $\mathbf{H}$ in the case of linear displacement, periodic and uniform traction boundary conditions can be found in \cite{Lukasz2008, ZUllah_CAS_2016} and is not repeated here. 

Matrix $\mathbf{K}$ comprises contributions of the matrix, yarns and yarn-matrix interface elements. $\mathbf{F}_{n+1}^{i}$ is a vector of residuals and is written as  
\begin{equation}\label{eq_residual}
\mathbf{F}_{n+1}^{i}=\left\{\begin{array}{l}
\mathbf{C}^{T}\boldsymbol{\lambda}_{n+1}^{i}-\mathbf{F}_{n+1}^{\textrm{int}\,\,i}\\
\mathbf{C}\,\,\,\mathbf{u}_{n+1}^{i}-\mathbf{D}\overline{\boldsymbol{\varepsilon}}_{n+1}
\end{array}\right\},   \qquad \mathbf{D}=\int_{\partial\Omega_{\mu}}\mathbf{HN}^{T}\mathbf{X}d\partial\Omega_{\mu},
 \end{equation}
where $\mathbf{X}$ is a matrix of spatial coordinates, evaluated at Gauss points during numerical integration of the surface integrals and is given as
\begin{equation}
\mathbf{X}=\frac{1}{2}\left[\begin{array}{cccccc}
2y_1 & 0 & 0 & y_2 & y_3 & 0\\
0 & 2y_2 & 0 & y_1 & 0 & y_3\\
0 & 0 & 2y_3 & 0 & y_1 & y_2
\end{array}\right].
\end{equation}
At Newton-Raphson iteration $i$, variable $\boldsymbol{\xi}=\mathbf{u}, \boldsymbol{\lambda}$ is calculated using $\boldsymbol{\xi}_{n+1}^i=\boldsymbol{\xi}_n + \sum_{m=1}^i \boldsymbol{\xi}_{n+1}^{m}$. In Equation (\ref{eq_residual}), $\mathbf{F}_{n+1}^{\textrm{int}\,\,i}$ is a vector of internal forces. Furthermore, $\mathbf{C}\mathbf{u}_{n+1}^{i}$ and $\mathbf{C}^{T}\boldsymbol{\lambda}_{n+1}^{i}$ are associated with the RVE boundary conditions and are written as 
\begin{equation}\label{eq_mat_Cu_and_DLam}
\mathbf{C}\mathbf{u}_{n+1}^{i}=\intop_{\Gamma}\mathbf{HN}^{T}\mathbf{u}_{n+1}^{h\, i}d\Gamma,   \qquad
\mathbf{C}^{T}\boldsymbol{\lambda}_{n+1}^{i}=\intop_{\Gamma}\mathbf{HN}^{T}\boldsymbol{\lambda}_{n+1}^{h\, i}d\Gamma,
 \end{equation}
where $\mathbf{u}^h$ and $\boldsymbol{\lambda}^h$  are displacements and Lagrange multipliers calculated at a Gauss point, i.e. $\boldsymbol{\xi}^h=\mathbf{u}^h, \boldsymbol{\lambda}^h=\mathbf{N}\boldsymbol{\xi}^{e\,i}_{n+1}$, where $\boldsymbol{\xi}^{e}$ is a matrix of displacements or Lagrange multipliers associated with element $e$. Finally, the homogenised stress for global increment $n+1$ is written as: 
\begin{equation}
\overline{\boldsymbol{\sigma}}_{n+1}=\frac{1}{V}\mathbf{D}^T\boldsymbol{\lambda}_{n+1}, 
\end{equation}
To compute the homogenised stiffness matrix $\overline{\mathbf{C}}$ at the end of global increment $n+1$, the converged matrix $\mathbf{K}$ is subjected to six different macro-strain perturbations of unit vector leading to six linear system of equations. This will give a set of homogenised stresses, i.e. 
\begin{equation}\label{Eq_C_stress}
\overline{\mathbf{C}}=\left[\begin{array}{cccccc}
\overline{\boldsymbol{\sigma}}^{1} & \overline{\boldsymbol{\sigma}}^{2} & \overline{\boldsymbol{\sigma}}^{3} & \overline{\boldsymbol{\sigma}}^{4} & \overline{\boldsymbol{\sigma}}^{5} & \overline{\boldsymbol{\sigma}}^{6}\end{array}\right], 
\end{equation}
where for example: 
\begin{equation}
\begin{array}{cc}
\overline{\boldsymbol{\sigma}}^{1}: & \textrm{for}\,\,\,\,\overline{\boldsymbol{\varepsilon}}=\left[\begin{array}{cccccc}
1 & 0 & 0 & 0 & 0 & 0\end{array}\right]^{T}\\
\overline{\boldsymbol{\sigma}}^{4}: & \textrm{for}\,\,\,\,\overline{\boldsymbol{\varepsilon}}=\left[\begin{array}{cccccc}
0 & 0 & 0 & 1 & 0 & 0\end{array}\right]^{T}\\
\end{array}.
\end{equation}
In each of the six cases, only the right-hand side of the system of Equations (\ref{eq_RVE_disc_Eqs}) changes, which is solved very efficiently as the left-hand side matrix is factorised only once.

\section{Calibration and validation of plasticity model}\label{sec_calib_valid}
Following \cite{Stier2015}, the plasticity model is first calibrated against the experimental results from \cite{Fiedler2001, Melro2013_partI, Stier2015} for epoxy resin subjected to tensile and compressive loading. A list of material parameters used in this case is shown in Table \ref{table_MatrixOnly_properties}, where $E$, $\nu$, $\sigma_{t_{o}}$, $\sigma_{t_{o}}$ and $\nu_{plas}$ are given in \cite{Stier2015, Melro2013_partI}. Moreover, the hardening parameters, i.e. $H_t$, $H_c$, $n_t$ and $n_c$ are determined from the numerical simulation based on the experimental stress-strain curves. The estimated parameters $H_t$, $H_c$ are the same as given in \cite{Stier2015} but in contrast, due to the use of hardening law as a function of internal kinematic variables leads to different $n_t$ and $n_c$ in our case. 

\begin{figure}[h!]
\begin{centering}
\includegraphics[scale=0.8]{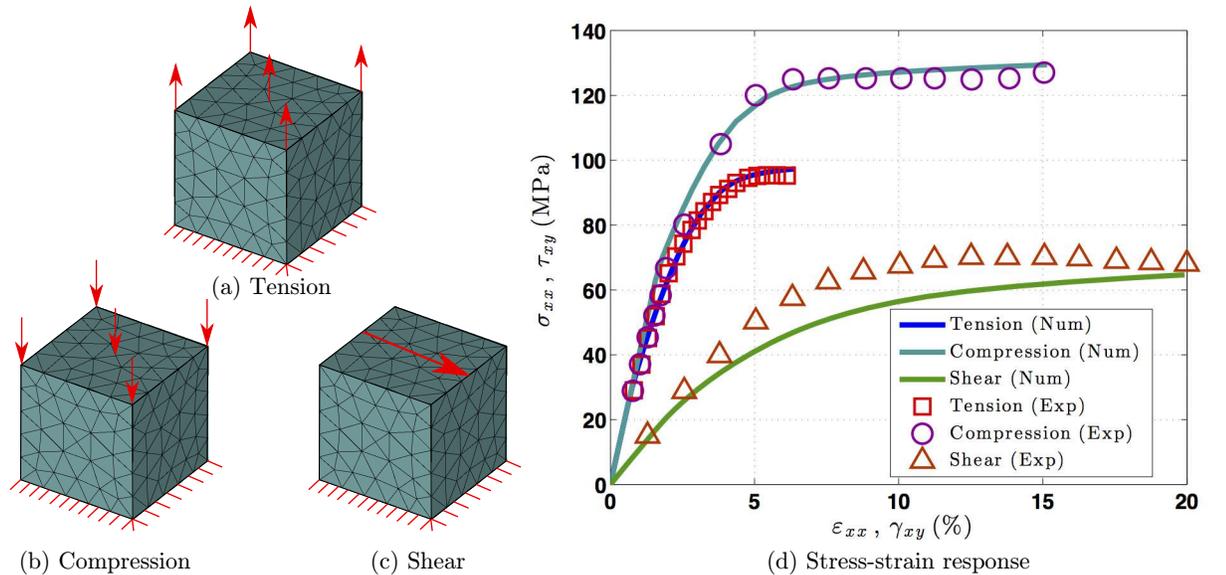}
\caption{Comparison of numerical and experimental \cite{Fiedler2001} stress-strain response for an epoxy resin} \label{Fig_MatrixOnlyMeshResults}
\end{centering}
\end{figure}

The geometry considered in this case is a cube of dimension 1mm, which is discretised with 1191 tetrahedral elements and 299 nodes. The cube is fixed at the bottom face and subjected to tension, compression and shear loading on the top face as shown in Figures \ref{Fig_MatrixOnlyMeshResults}(a), (b) and (c) respectively. A comparison between numerical and experimental stress-strain responses for all the three loading scenarios are shown in Figure \ref{Fig_MatrixOnlyMeshResults}(d). As expected the numerical and experimental responses in tension and compression are in good agreement, as a result of parameter fitting. The response in shear is not fitted and also shows fairly good agreement. All of the three responses are highly non-linear and it is clear that the plasticity model can capture them well. It must be noted that the plasticity model requires input data for tension and compression and can be used subsequently to simulate more generalised loading scenarios.
\begin{table}[h!]
\centering
\begin{tabular}{ll}
\hline
Parameter & Value \\ 
\hline
Young's modulus ($E$) & 3.76 GPa \\
Poisson's ratio ($\nu$) & 0.39 \\
Plastic Poisson's ratio ($\nu_{plas}$) & 0.3 \\ 
Initial yield strength in tension ($\sigma_{t_{o}}$) & 29 MPa\\ 
Initial yield strength in compression ($\sigma_{t_{o}}$) & 67 MPa\\ 
$H_t$ & 67 MPa\\ 
$H_c$ & 58 MPa\\ 
$n_t$ & 170 \\ 
$n_c$ & 150 \\ 
\hline
\end{tabular}
\caption{Material parameters for epoxy resin}
\label{table_MatrixOnly_properties}
\end{table}

\section{Numerical Examples}\label{sec_num_examples}
Three numerical examples are now given to demonstrate the correct implementation and performance of the developed computational framework.

\subsection{Unidirectional GFRP composites}\label{sec_Exp_UGFRP}
\begin{figure}[h!]
\begin{centering}
\includegraphics[scale=0.8]{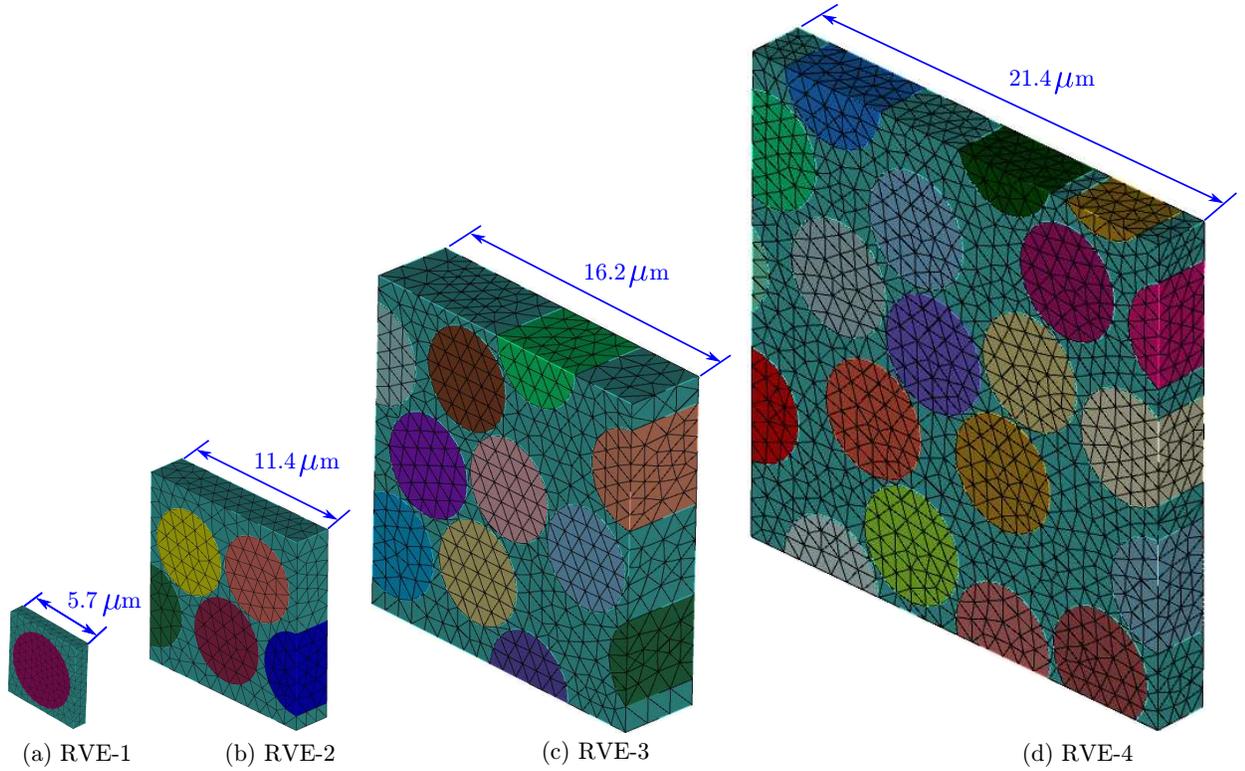}
\caption{Different RVE sizes for the UD GFRP example} \label{Fig_Example1_RVEs}
\end{centering}
\end{figure}

\begin{figure}[h!]
\begin{centering}
\includegraphics[scale=0.8]{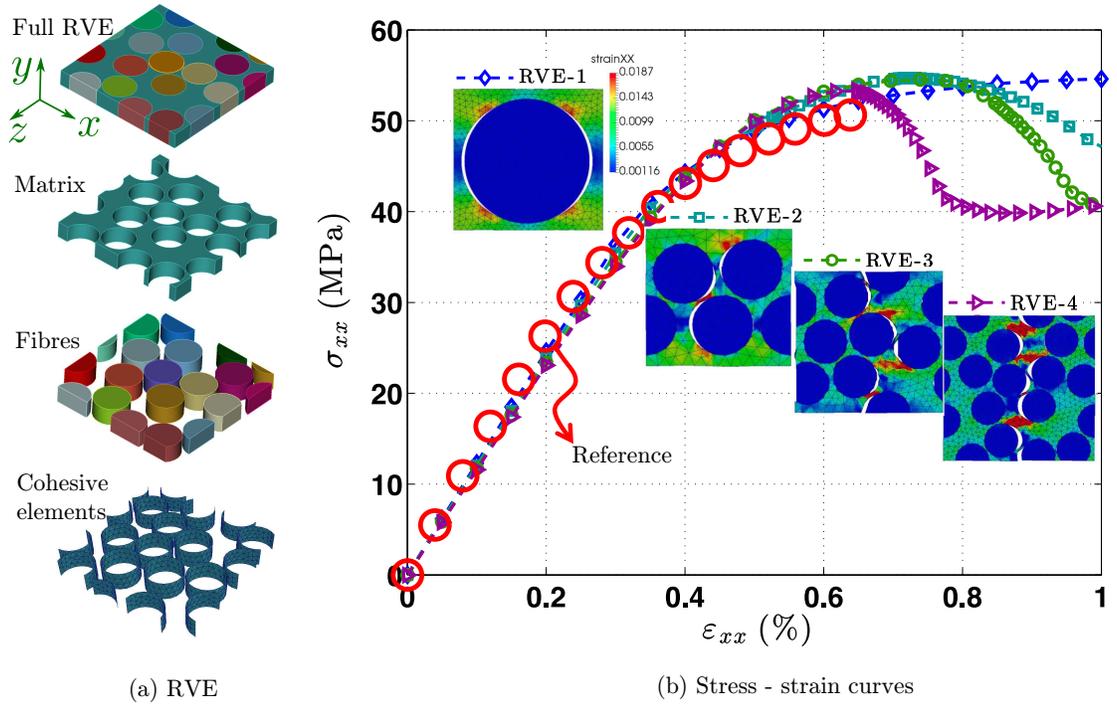}
\caption{RVE components and comparison of numerical and reference stress-strain curves for different RVE sizes for the UD GFRP example} \label{Fig_Example1_stress_strain_diff_RVE}
\end{centering}
\end{figure}

The first numerical example consists of polymer composites reinforced with unidirectional glass fibres subjected to transverse tension. A similar numerical example is also considered in \cite{Melro2013_partII}. Four different RVE sizes are considered in this case, consisting of periodic, randomly distributed but axially aligned fibres and are shown in Figure \ref{Fig_Example1_RVEs}. The algorithm proposed in \cite{Melro2008_rado_fibrer} is used to randomly generate the fibres within the RVEs with diameter of 5$\mu$m and volume fraction of 60\%.  The four RVEs are discretised with 1,499,  3,239, 13,023 and 21,140 tetrahedral elements. The detailed geometry for RVE-4, showing the individual matrix, fibres and cohesive interface elements, is presented in Figure \ref{Fig_Example1_stress_strain_diff_RVE}(a).

A list of material properties used for the elasto-plastic matrix materials are the same as given in Table \ref{table_MatrixOnly_properties}, while for the linear-elastic and isotropic glass fibres Young's modulus and Poisson's ratio used are 74 GPa and 0.2 respectively. For the cohesive interface elements, interface strength and fracture energy are 50 MPa and 2 J/$m^2$ respectively. The macro-strain (applied to the RVEs using periodic boundary conditions) versus homogenised stress response for all four RVEs are compared to reference numerical results from \cite{Melro2013_partII} and are shown in Figure \ref{Fig_Example1_stress_strain_diff_RVE}(b). The RVEs are subjected to a transverse strain $\varepsilon_{xx}$ of 1 percent. It is clear from Figure \ref{Fig_Example1_stress_strain_diff_RVE}(b) that the developed computational framework accurately predicts the stress-strain behaviour in the pre-peak region (up to $\varepsilon_{xx}=0.65\%$) for all of RVEs. The size effect can be clearly seen in the post-peak region (beyond $\varepsilon_{xx}=0.65\%$); increasing the size of the RVE leads to a more brittle response. A similar behaviour was also reported in \cite{Melro2013_partII, PhuNguyen2010}. Issues related with the existence and size of the RVE and pre- and post-peak region behaviour are described in detail in \cite{PhuNguyen2010, Gitman2007, Bazant2010, Ph_MM_eview_2011}, where the ill-posedness of the macro-level BVP and its non-objectivity with respect to the size of the RVE is discussed. A detailed description of dealing with these limitation of the classical CH schemes is given in \cite{Ph_MM_eview_2011}. These specialised treatments are not considered in this paper. The final damaged RVEs with clear localisation zones/debonding are also shown in Figure \ref{Fig_Example1_stress_strain_diff_RVE}(b). The damaged zones consist of fully damaged cohesive interface elements that are perpendicular to the direction of the applied strain. It is clear that fibre-matrix decohesion interface leads to a reduction in load transfer from the matrix to the fibres, which results in the overall stiffness. Furthermore, strain localisation associated with the damaged zones subsequently leads to severe plastic deformation of matrix material. 

\begin{figure}[h!]
\begin{centering}
\includegraphics[scale=0.8]{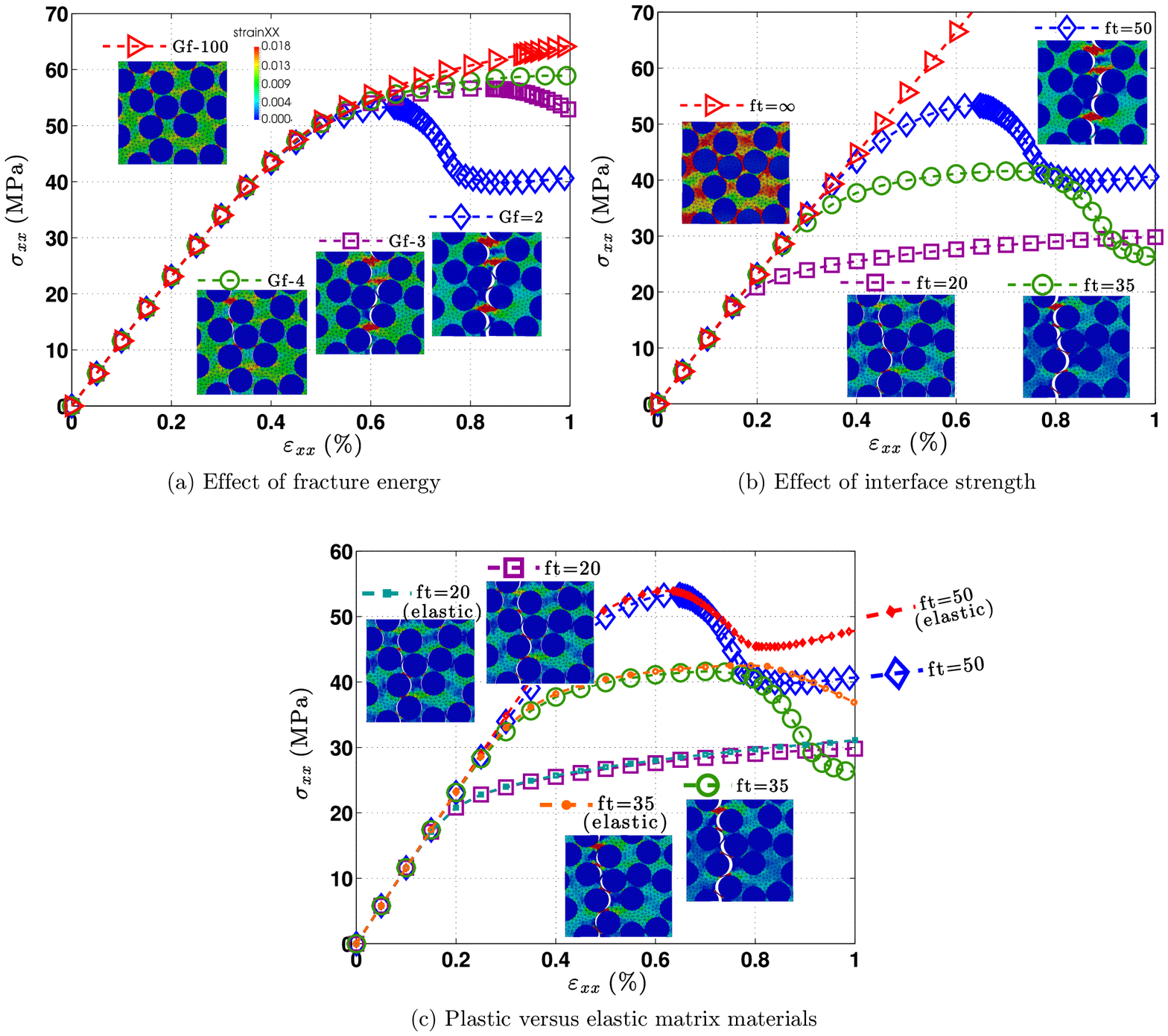}
\caption{Parametric study for the UD GFRP example} \label{Fig_Example1_paramaric}
\end{centering}
\end{figure}

A parametric study is also conducted to investigate the effect of different parameters on the macro-strain versus homogenised stress response. The effect of fracture energy on the stress-strain response is shown in Figure \ref{Fig_Example1_paramaric}(a), where fracture energies of 2, 3, 4 and 100 J/m$^2$ are used but all other parameters are kept constant. It is clear from Figure \ref{Fig_Example1_paramaric}(a) that the stress-strain response are the same for all the four cases up to $\varepsilon_{xx}=0.6\%$. Furthermore, lower fracture energies leads to clear damaged zone with high strain localisation (severe plastic deformation). The effect of interface strength on the stress-strain response is shown in Figure \ref{Fig_Example1_paramaric}(b), where interface strength of 20, 35 and 50 MPa are used and all other parameters are kept constant. Furthermore, the effect of considering unlimited interface strength, i.e. $f_t = \infty$ in shown in Figure  \ref{Fig_Example1_paramaric}(b). A clear localisation can be seen for both $f_t = 35$ and $f_t = 50$ cases as compared to the $f_t = 20$ for the applied strain of 1 \%. 

The effect of a linear elastic material, as opposed to an elasto-plastic material, on the strain-stress response is shown in Figure \ref{Fig_Example1_paramaric}(c). In addition, three cases with different interface strengths of cohesive interface elements, i.e. 20, 35 and 50 MPa are considered. It is clear from Figure \ref{Fig_Example1_paramaric}(c) that in the pre-peak regions, the use of either linear-elastic or elasto-plastic matrix material leads to almost similar stress-strain response while in the post-peak region the use of linear-elastic matrix material leads to relatively stiff response. The final damaged RVEs for both $f_t = 20$ and $f_t = 35$ with both linear-elastic and elasto-plastic matrix materials are also shown in Figure \ref{Fig_Example1_paramaric}(c). The high strain localisation in the damaged zones leads to severe plastic deformations leading to a more brittle stress-strain response. 

\subsection{Multi-fibres multi-layer RVE}\label{sec_Exp_M2RVE}
\begin{figure}[h!]
\begin{centering}
\includegraphics[scale=0.8]{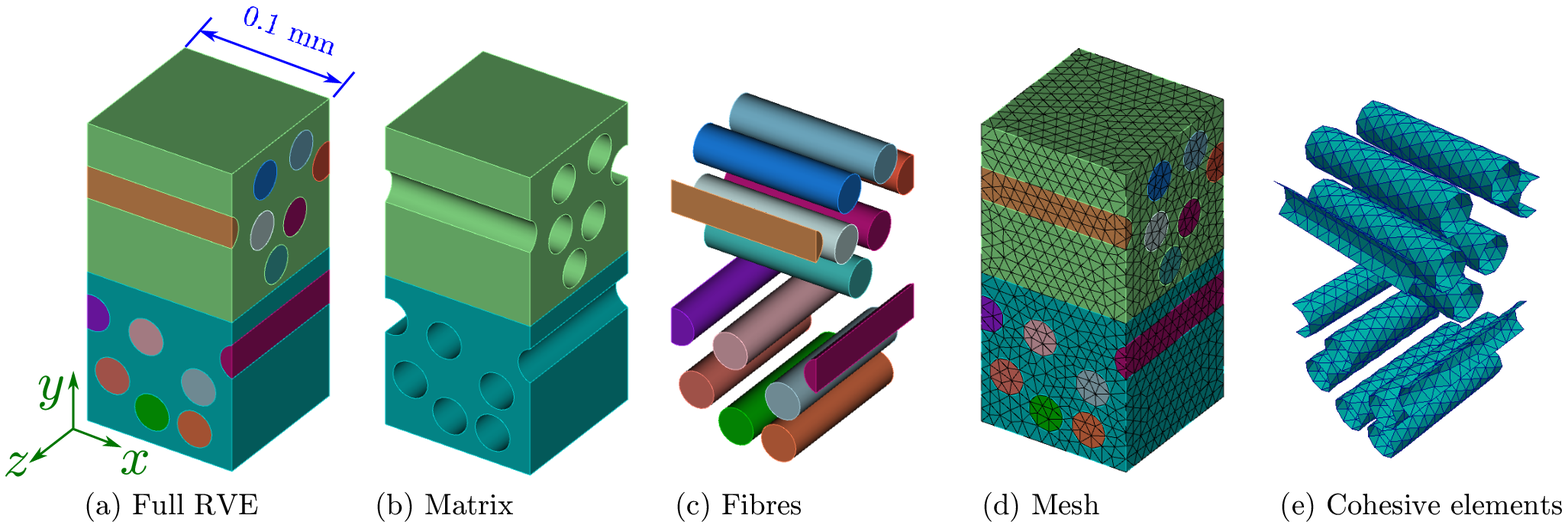}
\caption{Geometry and mesh for the M$^2$RVE example} \label{Fig_Example2_RVE}
\end{centering}
\end{figure}

A multi-fibre multi-layer RVE subjected to in-plane shear is considered in the second example. A similar example is also analysed experimentally and numerically in \cite{Soni2014_M2RVE}, the stress-strain response from which is used here as a reference. The UD FRP composite used in this case, consist of E-glass (ER-459L) and epoxy matrix (EPOFINE-556) with FINEHARD- 951 hardeners. The RVE geometry is shown in Figure \ref{Fig_Example2_RVE}(a), consisting of two cubes of dimension 1mm with randomly distributed fibres (generated using the algorithm in \cite{Melro2008_rado_fibrer}) of 24 $\mu$m and volume fraction of 28 \% and are placed on the top of each other with an angle of 90$^o$. In Figures \ref{Fig_Example2_RVE}(b) and (c) individual matrix and fibres are shown respectively. The RVE is discretised with 32,818 tetrahedral elements and is shown in Figure \ref{Fig_Example2_RVE}(d), while fibre-matrix interfaces are discretised with 3,056 cohesive interface elements and are shown in Figure \ref{Fig_Example2_RVE}(e). Moreover, a perfect bond is assumed between laminae. 

\begin{figure}[h!]
\begin{centering}
\includegraphics[scale=0.8]{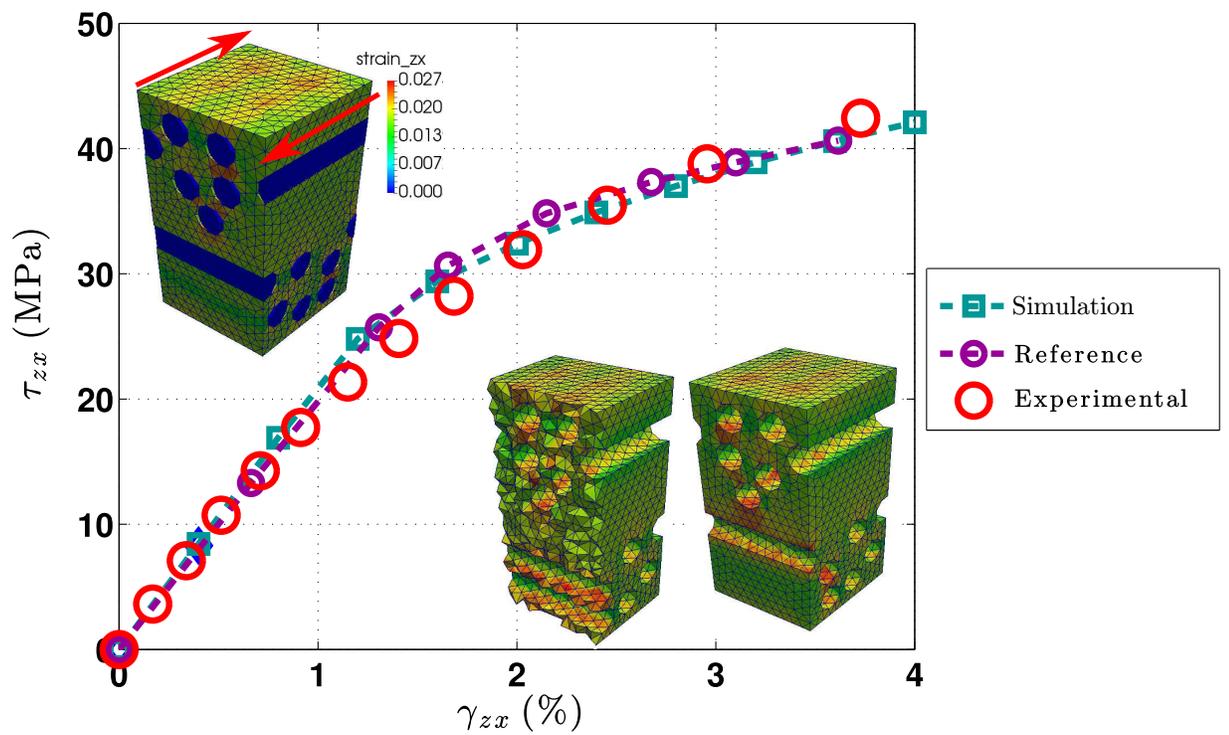}
\caption{Comparison of numerical and reference stress-strain response for the M$^2$RVE example} \label{Fig_Example2_stress_strain}
\end{centering}
\end{figure}

For the linear-elastic and isotropic glass fibres, Young's modulus and Poisson's ratio are 73 GPa and 0.23 respectively. For the matrix, most of the material parameters used are the same as given in Table \ref{table_MatrixOnly_properties} with the only exception of Young's modulus and Poisson's ratio, which are 4.7 GPa and 0.3 respectively. For cohesive interface elements, interface strength and fracture energy used are 30 MPa and 100 J/m$^2$ respectively \cite{Soni2014_M2RVE}. The RVE in this example is subjected to shear strain $\gamma_{zx}$ = 4\%, as shown in Figure \ref{Fig_Example2_stress_strain}. The shear stress versus shear strain ($\gamma_{zx}$ versus $\tau_{zx}$) response is compared with the experimental and numerical results from \cite{Soni2014_M2RVE} and is shown in Figure \ref{Fig_Example2_stress_strain}, which are in a very good agreement. Stress-strain response is almost linear up to $\gamma_{zx}$ = 1.5\%, beyond which the response is nonlinear due to the decohesion between fibres and matrix. The difference between the numerical and experimental results (especially between $\gamma_{zx}$ = 1\% and 2.5\%), might be due to the assumption of perfect bonding between the $0^o$ and $90^o$ laminae. At the end of the simulation, contours of $\gamma_{zx}$ over the deformed RVE are also shown in Figure \ref{Fig_Example2_stress_strain}. Strain is very small in the glass fibres as compared to matrix material due to the associated high stiffness. Decohesion between matrix and fibres can also be seen Figure \ref{Fig_Example2_stress_strain}.

\subsection{Plain weave textile composites}\label{sec_Exp_textile}
\begin{figure}[h!]
\begin{centering}
\includegraphics[scale=0.8]{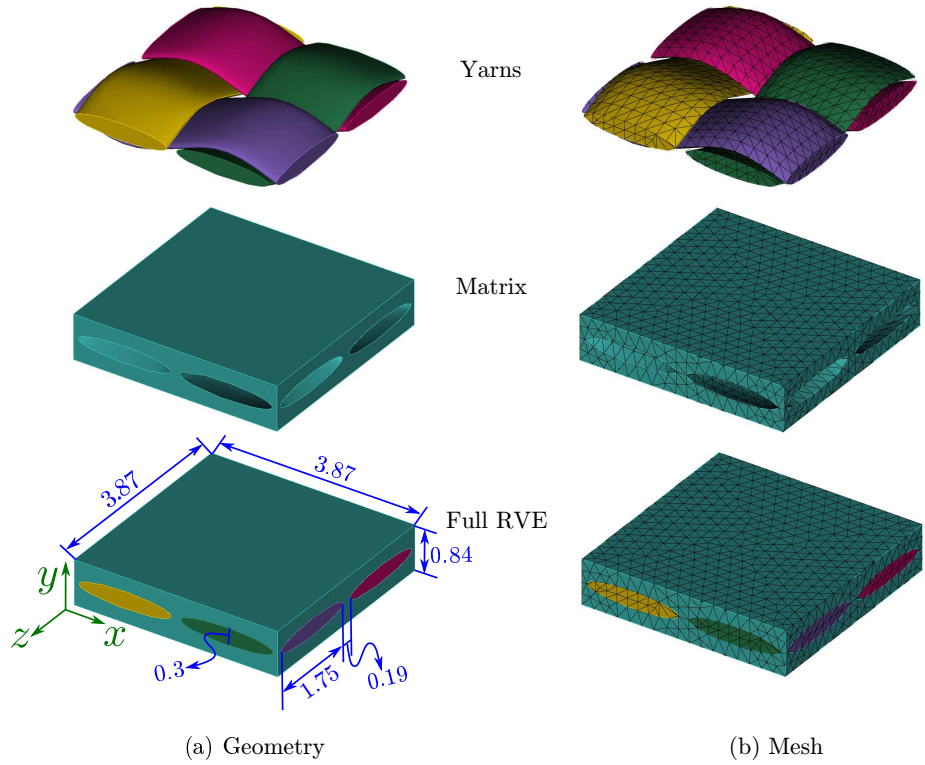}
\caption{Geometry and mesh for the plain weave textile composites example (dimensions in mm)} \label{Fig_Example3_RVE}
\end{centering}
\end{figure}

Finally, a plain weave textile composite subjected to a variety of normal and shear loading conditions is considered, consisting of E-glass fibres and epoxy matrix. A similar numerical example is also considered in \cite{kollegal2001_textile}. An RVE, consisting of similar yarns in warp and weft directions is used in this example, for which the geometry with all of the required dimensions are shown in Figure \ref{Fig_Example3_RVE}(a). Elliptical cross-sections and cubic splines are used respectively to model the cross sections and paths of the yarns. The volume fraction of fibres within the yarns is 65\% while the total volume fraction of fibres within the RVE is 35\%. The RVE is discretised with 11,516 tetrahedral elements and is shown in Figure  \ref{Fig_Example3_RVE}(b). For the elasto-plastic matrix material, the same properties are used as given in Table \ref{table_MatrixOnly_properties}, while for the linear-elastic and transversely isotropic yarns, material properties are given in Table \ref{table_YarnMaterial_properties} \cite{kollegal2001_textile}. Furthermore, a perfect bond is assumed between yarns and matrix. 

\begin{figure}[h!]
\begin{centering}
\includegraphics[scale=0.8]{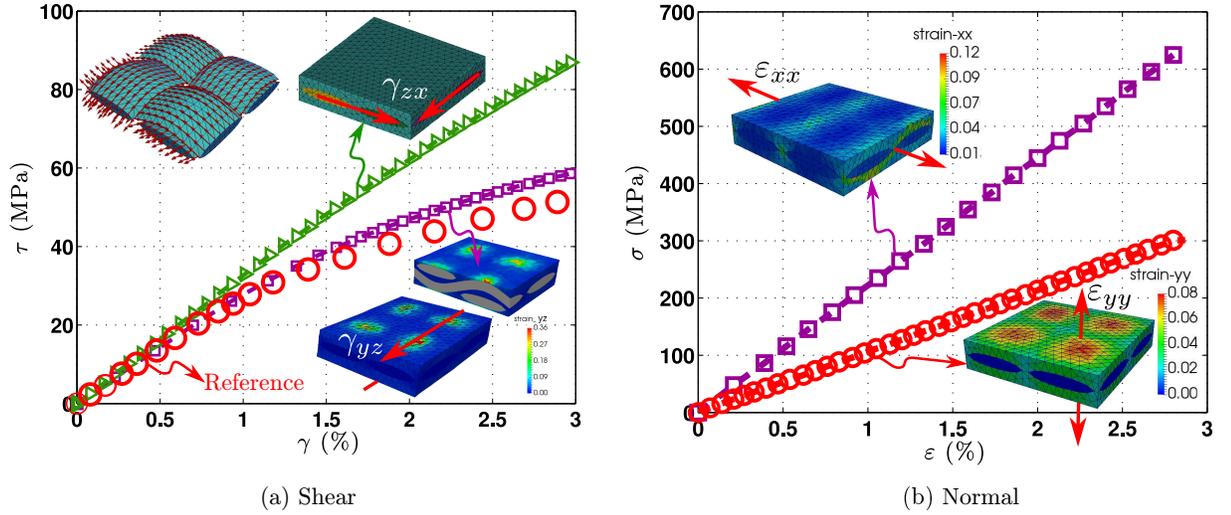}
\caption{Stress-strain responses for the plain weave textile composites subjected to different loading conditions} \label{Fig_Example3_stress_strain}
\end{centering}
\end{figure}

The yarns direction, calculated from the potential flow analysis, are shown in Figure \ref{Fig_Example3_stress_strain}(a). Four loading conditions, including two normal ($\varepsilon_{xx}$ and $\varepsilon_{yy}$) and two shear ($\gamma_{yz}$ and $\gamma_{zx}$) are considered, where the RVE is subjected to a macro-strain of 3\%. For the shear case, the stress-strain responses are shown in Figure \ref{Fig_Example3_stress_strain}(a). The nonlinear response beyond a strain of 1.5\% is due to matrix failure. The stress-strain response in the case of $\gamma_{yz}$ is also compared with numerical results from \cite{kollegal2001_textile}, which are in a very good agreement, especially in the linear region (up to $\gamma_{yz}$=1.5\%). Beyond $\gamma_{yz}$=1.5\%, our simulation result is relatively stiffer, which might be due to the use of linear-elastic material for the yarns. A high strain gradient can also be seen in the matrix, especially in regions of the thin matrix layer. The response in the case of $\gamma_{zx}$ involves shearing of yarns leading to stiffer behaviour as compared to $\gamma_{yz}$. Furthermore, response in the case of $\varepsilon_{xx}$ and $\varepsilon_{yy}$ are shown in Figure \ref{Fig_Example3_stress_strain}(b). For the given range of applied strains, stress-strain responses for both $\varepsilon_{xx}$ and $\varepsilon_{yy}$ are linear. Moreover, $\varepsilon_{xx}$ involves direct tensile load on the yarns and behave stiffer as compared to $\varepsilon_{yy}$ case, where strain is applied directly on the matrix. 
 
\begin{table}[h!]
\centering
\begin{tabular}{ll}
\hline
Parameter & Value \\ 
\hline
$E_p$       &  18.06 GPa\\ 
$\nu_p$      &  0.34 \\ 
$E_z$         & 48.47 GPa \\ 
$\nu_{pz}$  & 0.25 \\ 
$G_{zp}$    & 5.58 GPa \\ 
\hline
\end{tabular}
\caption{Yarns material parameters for the plain weave textile composites example}
\label{table_YarnMaterial_properties}
\end{table}

\section{Concluding remarks}\label{sec_conc_remarks}
A three-dimensional, nonlinear micro/meso-mechanical multi-scale CH framework is developed for FRP composites. The matrix material is modelled as elasto-plastic, using a paraboloidal yield surface. Decohsion of the fibre-matrix interface are modelled using cohesive interface elements. The yarns/fibres are modelled as linear-elastic and transversely isotropic material. It is shown that the two dominant damage mechanisms, i.e. matrix plasticity and fibre-matrix interfacial decohesion control the strength of FRP composites. Experimental stress-strain results for epoxy resin for both tension and compression load cases are used to calibrate the plasticity model and is validated subsequently for the shear loading. Three numerical examples with a variety of RVEs and loading conditions are considered to demonstrate the performance of the proposed computational framework. In both UD FRP composite and M$^2$RVE examples, fibres are randomly generated within the RVEs using a statistically proven random distribution algorithm. In the UD FRP numerical example, the developed computational framework can accurately predict the stress-strain behaviour in the pre-peak region, while in the post-peak region size dependent response is observed, which is natural in the case of first-order computational homogenisation. A parametric study is also conducted for the UD FRP numerical example, i.e. the effect of different matrix and fibre-matrix interface parameters on the stress-strain behaviour and it is shown that failure starts at fibre-matrix interface followed by formation of a shear band or matrix plasticity. Furthermore, from the M$^2$RVE and plain weave textile composite numerical examples, it is shown that the computational framework can accurately predict the stress-strain behaviour of RVEs with complicated geometries subjected to different loading scenarios. The developed computational framework is implemented in the authors' open-source FE software MOFEM; this has additional capabilities, including generalised RVE boundary conditions, hierarchic finite elements and optimisation for high-performance computing. The developed computational framework provides the nonlinear micro/meso-mechanical response at lamina level, which will be used subsequently to simulate FRP composites at both laminate and structure level.  

\section*{Acknowledgements}
\begin{itemize}
\item The authors gratefully acknowledge the support of the UK Engineering and Physical Sciences Research Council through the Providing Confidence in Durable Composites (DURACOMP) project (Grant Ref.: EP/K026925/1).
\item The first author also acknowledges A. R. Melro (University of Bristol, UK) for providing the MATLAB code for generation of randomly distributed fibres within the RVE for the UD FRP composites. 
\end{itemize}

\section*{References}
\bibliography{bibfile}

\begin{thebibliography}{10}
\expandafter\ifx\csname url\endcsname\relax
  \def\url#1{\texttt{#1}}\fi
\expandafter\ifx\csname urlprefix\endcsname\relax\def\urlprefix{URL }\fi
\expandafter\ifx\csname href\endcsname\relax
  \def\href#1#2{#2} \def\path#1{#1}\fi

\bibitem{Tong2002}
L.~Tong, A.~P. Mouritz, M.~K. Bannister, 3D Fibre Reinforced Polymer
  Composites, Elsevier Science Ltd, The Boulevard, Langford Lane, Kidlington
  Oxford, UK, 2002.

\bibitem{Mouritz1999}
A.~Mouritz, M.~Bannister, P.~Falzon, K.~Leong, Review of applications for
  advanced three-dimensional fibre textile composites, Composites Part A:
  Applied Science and Manufacturing 30~(12) (1999) 1445 -- 1461.

\bibitem{ZUllah_CAS_2016}
Z.~Ullah, L.~Kaczmarczyk, S.~A. Grammatikos, M.~C. Evernden, C.~J. Pearce.,
  Multi-scale computational homogenisation to predict the long-term durability
  of composite structures, Computers and Structures (Under Review).

\bibitem{NematNaseer1993}
S.~Nemat-Nasser, M.~Hori, Micromechanics: Overall Properties of Heterogeneous
  Materials, Elsevier Science Publishers, North Holland, Amsterdam, 1993.

\bibitem{Geers2010}
M.~Geers, V.~Kouznetsova, W.~Brekelmans, Multi-scale computational
  homogenization: Trends and challenges, Journal of Computational and Applied
  Mathematics 234~(7) (2010) 2175 -- 2182.

\bibitem{Xiaoyi2016_textile}
X.-Y. Zhou, P.~D. Gosling, C.~J. Pearce, Z.~Ullah, L.~Kaczmarczyk,
  Perturbation-based stochastic multi-scale computational homogenization method
  for woven textile composites, International Journal of Solids and Structures
  80 (2016) 368 -- 380.

\bibitem{UllahACME2014}
Z.~Ullah, L.~Kaczmarczyk, M.~Cortis, C.~J. Pearce, Multiscale modelling of the
  textile composite materials, in: 22$^{nd}$ UK Conference of the Association
  for Computational Mechanics in Engineering (ACME), University of Exeter,
  Exeter, UK, 2014, pp. 214--217.

\bibitem{UllahACME2015}
Z.~Ullah, L.~Kaczmarczyk, C.~J. Pearce, Multiscale computational homogenisation
  to predict the long-term durability of composite structures, in: 23$^{rd}$ UK
  Conference of the Association for Computational Mechanics in Engineering
  (ACME), University of Swansea, Swansea, UK, 2015, pp. 95--98.

\bibitem{UllahACME2016}
Z.~Ullah, L.~Kaczmarczyk, C.~J. Pearce, Nonlinear micro-mechanical response of
  the fibre-reinforced polymer composites including matrix damage and
  fibre-matrix decohesion, in: 24$^{th}$ UK Conference of the Association for
  Computational Mechanics in Engineering (ACME), University of Cardiff,
  Cardiff, UK, 2016, pp. 256--259.

\bibitem{Gonzalez2007}
C.~Gonz\'alez, J.~LLorca, Mechanical behavior of unidirectional
  fiber-reinforced polymers under transverse compression: Microscopic
  mechanisms and modeling, Composites Science and Technology 67~(13) (2007)
  2795--2806.

\bibitem{NayaLLorca2015_incollection}
F.~Naya, C.~Lopes, C.~Gonz{\'a}lez, J.~LLorca, Computational micromechanics
  strategies for the analysis of failure in unidirectional composites, in:
  P.~P. Camanho, S.~R. Hallett (Eds.), Numerical Modelling of Failure in
  Advanced Composite Materials, Woodhead Publishing Series in Composites
  Science and Engineering, Woodhead Publishing, 2015, pp. 411 -- 433.

\bibitem{Brockenbrough1991}
J.~Brockenbrough, S.~Suresh, H.~Wienecke, Deformation of metal-matrix
  composites with continuous fibers: geometrical effects of fiber distribution
  and shape, Acta Metallurgica et Materialia 39~(5) (1991) 735 -- 752.

\bibitem{Vaughan2011}
T.~Vaughan, C.~McCarthy, Micromechanical modelling of the transverse damage
  behaviour in fibre reinforced composites, Composites Science and Technology
  71~(3) (2011) 388 -- 396.

\bibitem{Vaughan2010}
T.~Vaughan, C.~McCarthy, A combined experimental--numerical approach for
  generating statistically equivalent fibre distributions for high strength
  laminated composite materials, Composites Science and Technology 70~(2)
  (2010) 291 -- 297.

\bibitem{Soni2014_M2RVE}
G.~Soni, R.~Singh, M.~Mitra, B.~G. Falzon, Modelling matrix damage and
  fibre-matrix interfacial decohesion in composite laminates via a multi-fibre
  multi-layer representative volume element ({M}$^2${RVE}), International
  Journal of Solids and Structures 51~(2) (2014) 449 -- 461.

\bibitem{Digimat2011}
Digimat Users' Manual, Version 4.2.1 (2011).

\bibitem{Romanowicz2012}
M.~Romanowicz, A numerical approach for predicting the failure locus of fiber
  reinforced composites under combined transverse compression and axial
  tension, Computational Materials Science 51~(1) (2012) 7 -- 12.

\bibitem{Parambil2015}
N.~K. Parambil, S.~Gururaja, Micromechanical damage analysis in laminated
  composites with randomly distributed fibers, Journal of Composite Materials
  -~(-) (2015) 1--14.

\bibitem{Melro2013_partI}
A.~R. Melro, P.~P. Camanho, F.~M.~A. Pires, S.~T. Pinho, Micromechanical
  analysis of polymer composites reinforced by unidirectional fibres: Part {I}
  -- constitutive modelling, International Journal of Solids and Structures
  50~(11--12) (2013) 1897 -- 1905.

\bibitem{Melro2013_partII}
A.~R. Melro, P.~P. Camanho, F.~M.~A. Pires, S.~T. Pinho, Micromechanical
  analysis of polymer composites reinforced by unidirectional fibres: Part {II}
  -- micromechanical analyses, International Journal of Solids and Structures
  50~(11--12) (2013) 1906 -- 1915.

\bibitem{Tschoegl1971}
N.~Tschoegl, Failure surfaces in principal stress space, Journal of polymer
  science Part C: Polymer symposia 32~(1) (1971) 239--267.

\bibitem{Melro2008_rado_fibrer}
A.~R. Melro, P.~P. Camanho, S.~T. Pinho, Generation of random distribution of
  fibres in long-fibre reinforced composites, Composites Science and Technology
  68~(9) (2008) 2092 -- 2102.

\bibitem{Matveev_Long_2015_incollection}
M.~Matveev, A.~Long, Numerical modelling for predicting failure in textile
  composites, in: P.~P. Camanho, S.~R. Hallett (Eds.), Numerical Modelling of
  Failure in Advanced Composite Materials, Woodhead Publishing Series in
  Composites Science and Engineering, Woodhead Publishing, 2015, pp. 435 --
  455.

\bibitem{kollegal2001_textile}
M.~Kollegal, S.~Chatterjee, G.~Flanagan, Progressive failure analysis of plain
  weaves using damage mechanics based constitutive laws, International journal
  of damage mechanics 10~(4) (2001) 301--323.

\bibitem{ZakoTextileCDM2003}
M.~Zako, Y.~Uetsuji, T.~Kurashiki, Finite element analysis of damaged woven
  fabric composite materials, Composites Science and Technology 63~(3--4)
  (2003) 507 -- 516.

\bibitem{Blackketter1993}
D.~Blackketter, D.~Walrath, A.~Hansen, Modeling damage in a plain weave
  fabric-reinforced composite material, Journal of Composites, Technology and
  Research 15~(2) (1993) 136--142.

\bibitem{Ivanov2009}
D.~S. Ivanov, F.~Baudry, B.~V.~D. Broucke, S.~V. Lomov, H.~Xie, I.~Verpoest,
  Failure analysis of triaxial braided composite, Composites Science and
  Technology 69~(9) (2009) 1372 -- 1380.

\bibitem{Falzon2011CDM_2011_PartI}
B.~G. Falzon, P.~Apruzzese, Numerical analysis of intralaminar failure
  mechanisms in composite structures. {Part I: FE} implementation, Composite
  Structures 93~(2) (2011) 1039--1046.

\bibitem{Falzon2011CDM_2011_PartII}
B.~G. Falzon, P.~Apruzzese, Numerical analysis of intralaminar failure
  mechanisms in composite structures. {Part II: Applications}, Composite
  Structures 93~(2) (2011) 1047--1053.

\bibitem{Stier2015}
B.~Stier, J.~W. Simon, S.~Reese, Comparing experimental results to a numerical
  meso-scale approach for woven fiber reinforced plastics, Composite Structures
  122 (2015) 553 -- 560.

\bibitem{MoFEM_2016}
Mesh Oriented Finite Element Method ({MoFEM}) Version 0.3.8, University of
  Glasgow, Glasgow, UK, http://mofem.eng.gla.ac.uk/mofem/html/ (2016).

\bibitem{whitcomb2002}
J.~Whitcomb, X.~Tang, Micromechanics of moisture diffusion in composites with
  impermeable fibers, Journal of Composite Materials 36~(9) (2002) 1093--1101.

\bibitem{hobbiebrunken2006}
T.~Hobbiebrunken, M.~Hojo, T.~Adachi, C.~De~Jong, B.~Fiedler, Evaluation of
  interfacial strength in {CF}/epoxies using {FEM} and in-situ experiments,
  Composites Part A: Applied Science and Manufacturing 37~(12) (2006)
  2248--2256.

\bibitem{Maligno2008}
A.~R. Maligno, N.~A. Warrior, A.~C. Long, Finite element investigations on the
  microstructure of fibre-reinforced composites, eXPRESS Polymer Letters 2~(9)
  (2008) 665--676.

\bibitem{Lukasz2008}
L.~Kaczmarczyk, C.~J. Pearce, N.~Bi\'cani\'c, Scale transition and enforcement
  of {RVE} boundary conditions in second-order computational homogenization,
  International Journal for Numerical Methods in Engineering 74~(3) (2008)
  506--522.

\bibitem{Cubit}
{CUBIT}, The Geometry and Mesh Generation Toolkit, Version 13.0, Sandia
  National Laboratories, USA, https://cubit.sandia.gov/.

\bibitem{Paraview}
{ParaView} , Open-source, multi-platform data analysis and visualization
  application, Version 4.4.0, http://www.paraview.org/.

\bibitem{Segurado2004_cohesiveInterface}
J.~Segurado, J.~LLorca, A new three-dimensional interface finite element to
  simulate fracture in composites, International Journal of Solids and
  Structures 41~(11--12) (2004) 2977 -- 2993.

\bibitem{Fiedler2001}
B.~Fiedler, M.~Hojo, S.~Ochiai, K.~Schulte, M.~Ando, Failure behavior of an
  epoxy matrix under different kinds of static loading, Composites Science and
  Technology 61~(11) (2001) 1615 -- 1624.

\bibitem{PhuNguyen2010}
V.~P. Nguyen, O.~Lloberas-Valls, M.~Stroeven, L.~J. Sluys, On the existence of
  representative volumes for softening quasi-brittle materials -- a failure
  zone averaging scheme, Computer Methods in Applied Mechanics and Engineering
  199~(45--48) (2010) 3028 -- 3038.

\bibitem{Gitman2007}
I.~Gitman, H.~Askes, L.~Sluys, Representative volume: Existence and size
  determination, Engineering Fracture Mechanics 74~(16) (2007) 2518 -- 2534.

\bibitem{Bazant2010}
Z.~P. Bazant, Can multiscale-multiphysics methods predict softening damage and
  structural failure?, International Journal for Multiscale Computational
  Engineering 8~(1) (2010) 61--67.

\bibitem{Ph_MM_eview_2011}
V.~P. Nguyen, M.~Stroeven, L.~J. Sluys, Multiscale continuous and discontinuous
  modeling of heterogeneous materials: A review on recent developments, Journal
  of Multiscale Modelling 03~(04) (2011) 229--270.

\end{thebibliography}

\end{document}